\documentclass[10pt]{article}
\usepackage{fancyhdr}
\usepackage{extramarks}
\usepackage{amsmath}
\usepackage{amsthm}
\usepackage{amsfonts}
\usepackage{siunitx}
\usepackage{tikz}
\usepackage[plain]{algorithm}
\usepackage{algpseudocode}
\usepackage{multirow}
\usepackage{booktabs}
\usepackage{palatino}
\usepackage{graphicx}
\usepackage{subfigure}
\usepackage[colorlinks,linkcolor=black,anchorcolor=black,citecolor=black,urlcolor=blue]{hyperref}
\usepackage{amsmath,bm}
\usepackage{booktabs}
\usepackage{mathtools}
\usepackage{amssymb}
\usepackage{caption}
\usepackage{capt-of}
\usepackage{mciteplus}
\usepackage{cite}
\usepackage{mathrsfs}
\usepackage[title,titletoc,toc]{appendix}
\usepackage{xr}
\usepackage{parskip}
\usepackage{soul}
\usepackage{textcomp}
\usepackage[colaction]{multicol}
\usepackage[switch]{lineno}
\usepackage{lipsum}
\usepackage{etoolbox}
\usepackage{longtable}
\usepackage{array}
\usepackage{tablefootnote}
\usepackage{ragged2e}
\newcolumntype{C}[1]{>{\centering\arraybackslash}p{#1}}
\usetikzlibrary{automata,positioning}
\usetikzlibrary{automata,positioning}

\begin{document}

\title{Emerging vaccine-breakthrough SARS-CoV-2 variants}
\author{Rui Wang$^1$, Jiahui Chen$^1$, Yuta Hozumi$^1$, Changchuan Yin$^2$, and Guo-Wei Wei$^{1,3,4}$\footnote{
		Corresponding author.		Email: weig@msu.edu} \\
$^1$ Department of Mathematics, \\
Michigan State University, MI 48824, USA.\\
$^2$ Department of Mathematics, Statistics, and Computer Science, \\
University of Illinois at Chicago, Chicago, IL 60607, USA\\
$^3$ Department of Electrical and Computer Engineering,\\
Michigan State University, MI 48824, USA. \\
$^4$ Department of Biochemistry and Molecular Biology,\\
Michigan State University, MI 48824, USA. \\
}
\date{\today} 

\maketitle

\begin{abstract}
The recent global surge in coronavirus disease 2019 (COVID-19) infections have been fueled by new severe acute respiratory syndrome coronavirus 2 (SARS-CoV-2) variants, namely Alpha, Beta, Gamma, Delta, etc. The molecular mechanism underlying such surge is elusive due to the existence of 28,554, including 4,653 non-degenerate mutations on the spike (S) protein, which is the target of most COVID-19 vaccines. The understanding of the molecular mechanism of SARS-CoV-2 transmission and evolution is a prerequisite to foresee the global trend of emerging vaccine-breakthrough SARS-CoV-2 variants and the design of mutation-proof vaccines and monoclonal antibodies (mAbs). We integrate the genotyping of 1,489,884 SARS-CoV-2 genomes isolated from patients, a library collection of 130 human antibodies, tens of thousands of mutational data points, topological data analysis (TDA), and deep learning to reveal SARS-CoV-2 evolution mechanism and forecast emerging vaccine-escape variants.    
We show that infectivity-strengthening and antibody-disruptive co-mutations on the S protein receptor-binding domain (RBD) can quantitatively explain the infectivity and virulence of all prevailing variants. We demonstrate that Lambda is as infectious as Delta but is more vaccine-resistant. We analyze emerging vaccine-breakthrough co-mutations in 20 COVID-19 devastated countries, including  the United Kingdom (UK), the United States (US), Denmark (DK), Brazil (BR), Germany (DE), Netherlands (NL), Sweden (SE), Italy (IT), Canada (CA), France (FR), India (IN), and Belgium (BE), etc. We envision that natural selection through infectivity will continue to be a main  mechanism for viral evolution among unvaccinated populations, while antibody disruptive co-mutations will fuel the future growth of vaccine-breakthrough variants among fully vaccinated populations. Finally, we have identified the following sets of co-mutations that have the great  likelihood of becoming dominant:  [A411S, L452R, T478K], [L452R, T478K, N501Y], [V401L, L452R, T478K],  [K417N, L452R, T478K], [L452R, T478K, E484K, N501Y],   and [P384L, K417N, E484K, N501Y]. We predict they, particularly the last four,  will break through existing vaccines. We foresee an urgent need to   develop  new vaccines that target these co-mutations. 


\end{abstract}
 \newpage

\setcounter{page}{1}
\renewcommand{\thepage}{{\arabic{page}}}


\section{Introduction}

The death toll of coronavirus disease 2019 (COVID-19) caused by severe acute respiratory syndrome coronavirus 2 (SARS-CoV-2) has exceeded 4.4 million in August 2021. Tremendous efforts in combating SARS-CoV-2 have led to several authorized vaccines,  
which mainly target the viral spike (S) proteins. However, the emergence of mutations on the S gene has resulted in more infectious variants and vaccine breakthrough infections.  Emerging vaccine breakthrough SARS-CoV-2 variants pose a grand challenge to the long-term control and prevention of the COVID-19 pandemic. Therefore, forecasting emerging breakthrough SARS-CoV-2 variants is of paramount importance for the design of new mutation-proof vaccines and monoclonal antibodies (mABs).


To predict emerging breakthrough SARS-CoV-2 variants, one must understand the molecular mechanism of viral transmission and evolution, which is one of the greatest challenges of our time. SARS-CoV-2 entry of a host cell depends on the binding between S protein and the host angiotensin-converting enzyme 2 (ACE2), primed by host transmembrane protease, serine 2 (TMPRSS2) \cite{hoffmann2020sars}. Such a process inaugurates the host's adaptive immune response, and consequently,  antibodies are generated to combat the invading virus either through direct neutralization or non-neutralizing binding \cite{chen2020review,chen2021sars}. 
S protein receptor-binding domain (RBD) is a short immunogenic fragment that facilitates the S protein binding with ACE2. Epidemiological and biochemical studies have suggested that the binding free energy (BFE) between the S RBD and the ACE2 is proportional to the infectivity \cite{li2005bats,qu2005identification,song2005cross,hoffmann2020sars,walls2020structure}. Additionally, the strong binding between the RBD and mAbs leads to effective direct neutralization   \cite{wang2020human,yu2020receptor,li2021impact}. Therefore, RBD mutations have dominating impacts on viral infectivity, mAb efficacy, and vaccine protection rates. Mutations may occur for various reasons, including random genetic drift,  replication error, polymerase error, host immune responses,   gene editing, and recombinations  \cite{sanjuan2016mechanisms,grubaugh2020making,kucukkal2015structural,yue2005loss,wang2020host}. Being beneficial from the genetic proofreading mechanism regulated by NSP12 (a.k.a RNA-dependent RNA polymerase) and NSP14 \cite{sevajol2014insights,ferron2018structural}, SARS-CoV-2 has a higher fidelity in its replication process than the other RNA viruses such as influenza. Nonetheless, near 700 non-degenerate mutations are observed on RBD, contributing many key mutations in emerging variants, i.e., N501Y for Alpha, K417N, E484K, and N501Y for Beta, K417T, E484K, and N501Y for Gamma, L452R and T478K for Delta, L452Q and F490S for Lambda, etc \cite{wang2021vaccine}. Given the importance of the RBD for SARS-CoV-2 infectivity, vaccine efficacy, and mAb effectiveness, it is imperative to understand the mechanism governing RBD mutations.  

In June 2020, when there were only 89 non-degenerated mutations on the RBD, and the highest observed mutational frequency was only around 50 globally, we were able to show that natural selection underpins SARS-CoV-2 evolution, based on the genotyping of 24,715 SARS-CoV-2 sequences isolated patients and a topology-based deep learning model for RBD-ACE2 binding analysis \cite{chen2020mutations}. In the same work, we predicted that RBD residues 452 and 501 ``have high chances to mutate into significantly more infectious COVID-19 strains'' \cite{chen2020mutations}. Currently, these residues are the key mutational sites of all prevailing SARS-CoV-2 variants.  We further foresaw a list of 1,149 most likely RBD mutations among 3686 possible RBD mutations \cite{chen2020mutations}. Up to date, every one of the observed 683 RBD mutations  belongs to the list. In April 2021, we demonstrated that all the 100 most observed RBD mutations of 651 existing RBD mutations from 506,768 viral genomes had enhanced the binding between RBD and ACE2, resulting in more infectious variants \cite{wang2021vaccine}. The odd for these 100 most observed mutations to be there  accidentally is smaller than one chance in 1.2 nonillions ($2^{100}\approx$ 1.2$\times 10^{30}$)\footnote{The average BFE changes of 1149 RBD mutations for the RBD-ACE2 complex is -0.28kcal/mol. Randomly, each RBD mutation has a 50\% chance to assume a BFE change above or below -0.28kcal/mol, which leads to $2^{100}=1.276506\times10^{30}$ possible states for 100 mutations.}. There is no double that natural selection via viral infectivity, rather than any other competing theories  \cite{sanjuan2016mechanisms,grubaugh2020making,kucukkal2015structural,yue2005loss,wang2020host}, is the dominating mechanism for SARS-CoV-2 transmission and evolution. This mechanistic discovery lays the foundation for forecasting future emerging SASR-CoV-2 variants.  
   
Understanding SARS-CoV-2 variant threats to current vaccines and mAbs is another urgent issue facing the scientific community \cite{chen2021prediction}. The World Health Organization (WHO) identified variants of concern (VOCs) and variants of interest (VOIs). The former describes variants that have an increment in the transmissibility and virulence, or adversely affect the effectiveness of vaccines, therapeutics, and diagnostics with clear clinical correlation evidence. The latter describes variants that carry genetic changes, which are predicted or known to reduce neutralization by antibodies generated against vaccination, the efficacy of treatments, and affect transmissibility, virulence, disease severity, immune escape, diagnostics, etc., which cause significant community transmission and suggest an emerging risk to the public. Currently, WHO listed four VOCs, i.e., variants B.1.1.7 (Alpha) \cite{davies2021estimated,wang2021increased,emary2021efficacy}, B.1.351 (Beta) \cite{wang2021increased,madhi2021efficacy}, P.1 (Gamma)\cite{wang2021increased}, and B.1.617.2 (Delta) \cite{deng2021transmission}), and four VOIs, i.e., variants B.1.525 (Eta)\cite{jangra2021sars},  B.1.526 (Iota) \cite{jangra2021sars,annavajhala2021novel}, B.1.617.1 (Kappa) \cite{greaney2021comprehensive},   C.37 (Lambda)\cite{kimura2021sars}, and B.1.621 (Mu) (A general introduction about the prevailing and emerging variants is given  in   Section S1 of the Supporting Information.). Our hypothesis is that the severity of variants to infectivity, vaccine efficacy, and mAbs effectiveness depends mainly on how the associated RBD mutations impact the binding with ACE2 and antibodies. Based on this hypothesis, we collected and analyzed a library of antibodies and unveiled that most of the RBD mutations would weaken the binding of S protein and antibodies and disrupt the efficacy and reliability of antibody therapies and vaccines \cite{chen2021prediction}. We predicted ``the urgent need to develop new mutation-resistant vaccines and antibodies and prepare for seasonal vaccination'' in early 2021 \cite{chen2021prediction}. We further identified vaccine-escape (i.e., vaccine-breakthrough) mutations and fast-growing mutations \cite{wang2021vaccine}. Our predictions of the threats from VOCs and VOIs were in great agreement with experimental data \cite{chen2021revealing}.   

The objective of this work is to forecast emerging SARS-CoV-2 variants that pose an imminent threat to combating COVID-19 and long-term public health. To this end, we carry out an RBD-specific analysis of SARS-CoV-2 co-mutations involving a wide variety of combinations of 683 unique single mutations on the RBD. We take a unique approach that integrates viral genotyping of 1,489,884 complete genome sequences isolated from patients, algebraic topology algorithms that won the worldwide competition in computer-aided drug discovery \cite{nguyen2019mathematical},  deep learning models trained with tens of thousands of mutational data points \cite{chen2021revealing,chen2021prediction}, and a library of 130 SARS-CoV-2 antibody structures. By analyzing the frequency, binding free energy (BFE) changes, and antibody disruption counts of RBD  co-mutations, we reveal that nine RBD co-mutation sets, namely [L452R, T478K], [L452Q, F490S], [E484K, N501Y], [F490S, N501Y], [S494P, N501Y], [K417T, E484K, N501Y], [K417N, L452R, T478K], [K417N, E484K, N501Y], and [P384L, K417N, E484K, N501Y], may strongly disrupt existing vaccines and mAbs with relatively high infectivity and transmissibility among the populations. 
We predict that low-frequency co-mutation sets  [A411S, L452R, T478K], [L452R, T478K, N501Y], [V401L, L452R, T478K],  and [L452R, T478K, E484K, N501Y] are on the path to become dangerous new variants. The associated new mutations, P384L,  V401L, and A411S, call for the new design of boosting vaccines and mAbs.

\section{Results}

\subsection{Vaccine-breakthrough S protein RBD mutations}

\begin{figure}[ht!]
	\centering
	\includegraphics[width = 1\textwidth]{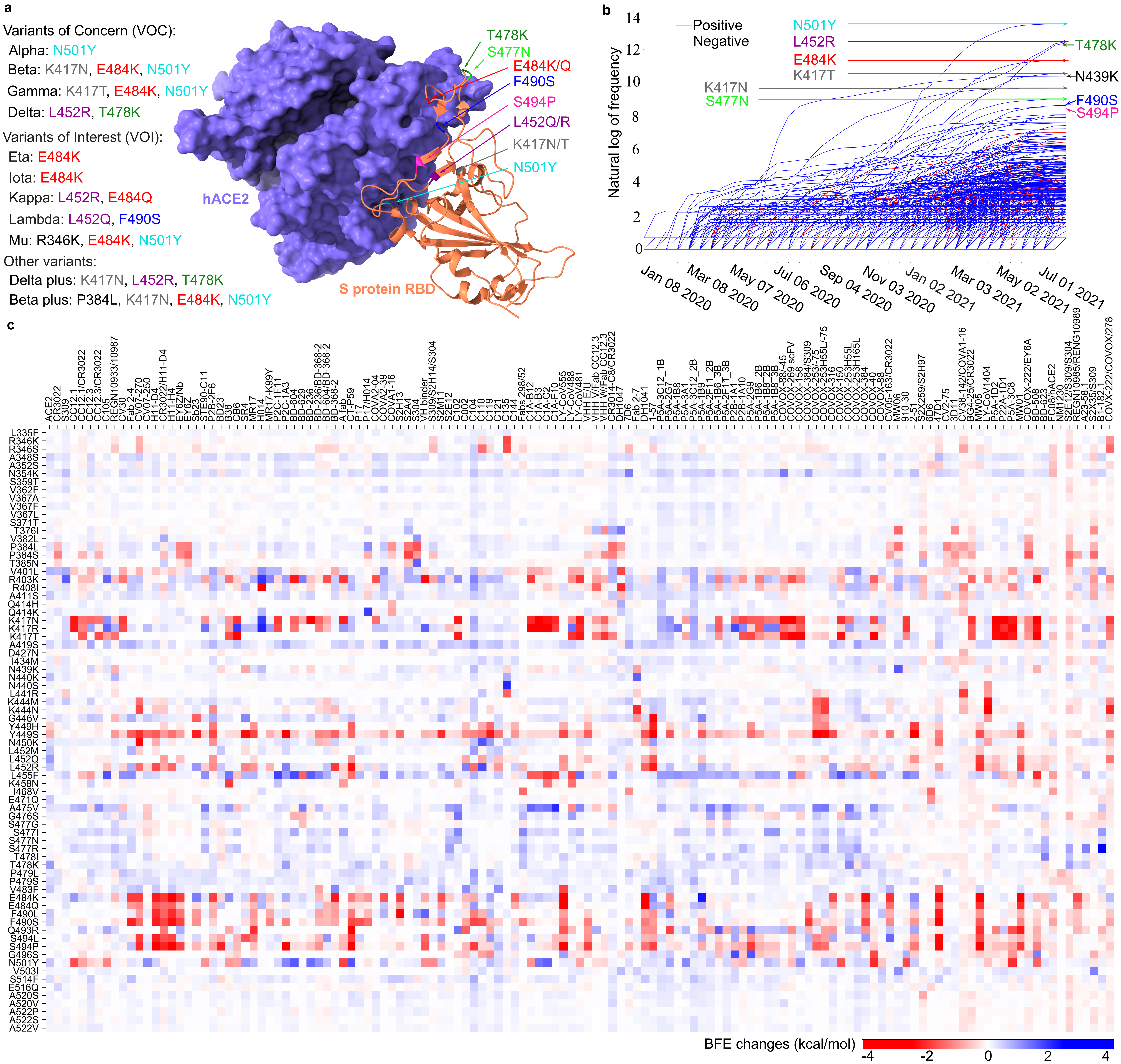}
	\caption{Most significant RBD mutations. 
	{\bf a} The 3D structure of SARS-CoV-2 S protein RBD and ACE2 complex (PDB ID: 6M0J). The RBD mutations in ten variants are marked with color. {\bf b} Illustration of the time evolution of 455 ACE2 binding-strengthening RBD mutations (blue) and 228 ACE2 binding-weakening RBD mutations (red). The $x$-axis represents the date and the $y$-axis represents the natural log of frequency. There has been a surge in the number of infections since early 2021. {\bf c} BFE changes of RBD complexes with ACE2 and 130 antibodies induced by 75 significant RBD mutations. A positive BFE change (blue) means the mutation strengthens the binding, while a negative BFE change (red) means the mutation weakens the binding.  Most mutations, except for vaccine-resistant Y449H and Y449S, strengthen the RBD binding with ACE2. Y449S and K417N are highly disruptive to antibodies. }
	\label{fig:6M0J}
\end{figure}
 
To understand the molecular mechanisms of vaccine-escape mutations, we 
analyze single nucleotide polymorphisms (SNPs) of 1,489,884 complete SARS-CoV-2 genome sequences, resulting in  683 non-degenerate RBD mutations and their associated frequencies. A full set of mutation information is available on our interactive web page \href{https://users.math.msu.edu/users/weig/SARS-CoV-2_Mutation_Tracker.html}{Mutation Tracker}. The infectivity of each mutation is mainly determined by the mutation-induced BFE change to the binding complex of RBD and ACE2. To estimate the impact of each mutation on vaccines, we collect a library of 130 antibody structures (Supporting Information S2.1.2), including Food and Drug Administration (FDA)-approved mAbs from  Eli Lilly and Regeneron. For a given RBD mutation, its number of antibody disruptions is given by the number of antibodies whose mutation-induced antibody-RBD BFE changes are smaller than -0.3kcal/mol (A list of names for antibodies that are disrupted by mutations can be found in the Supporting Information S2.1.1.). BFE changes following mutations are predicted by our deep learning model, TopNetTree \cite{wang2020topology}. We have created an interactive web page,  \href{https://weilab.math.msu.edu/MutationAnalyzer/}{Mutation Analyzer}, to list all RBD mutations, their observed frequencies, their RBD-ACE2 BFE changes following mutations, their number of antibody disruptions, and various ranks. Figure \ref{fig:6M0J}  illustrates RBD mutations associated with prevailing SARS-CoV-2 variants, time evolution trajectories of all RBD mutations, and the BFE changes of RBD-ACE2 and 130 RBD-antibodies induced by 75 significant mutations. A summary of our analysis is given in Table \ref{tab:WebMutaion}.

\begin{table}[ht!]
    \centering
    \setlength\tabcolsep{2pt}
	\captionsetup{margin=0.1cm}
	\caption{Top 25 most observed S protein RBD mutations. Here, BFE change refers to the BFE change for the S protein and human ACE2 complex induced by a single-site S protein RBD mutation. A positive mutation-induced BFE change strengthens the binding between S protein and ACE2, which results in more infectious variants. Counts of antibody disruption represent the number of antibody and S protein complexes disrupted by a specific RBD mutation.  Here, an antibody and S protein complex is to be disrupted if its binding affinity is reduced by more than 0.3 kcal/mol \cite{wang2021vaccine}. In addition, we calculate the antibody disruption ratio (\%), which is the ratio of the number of disrupted antibody and S protein complexes over 130 known complexes. Ranks are computed from 683 observed RBD mutations.}
    \label{tab:WebMutaion}
    {
    \begin{tabular}{c|cc|cc|ccc}
    \toprule
    \multirow{2}{*}{Mutation} & \multicolumn{2}{c|}{Worldwide} &  \multicolumn{2}{c|}{BFE change}  & \multicolumn{3}{c}{Antibody disruption} \\ \cline{2-8}  
     & Count  & Rank & Change & Rank & Count  & Ratio & Rank \\ 
    \midrule
    N501Y	&744354	&1	&0.5499	&30	&24	&18.46	&160 \\
    L452R	&259345	&2	&0.5752	&28	&39	&30.0	&98\\
    T478K	&239619	&3	&0.9994	&2	&2	&1.54	&557\\
    E484K	&84167	&4	&0.0946	&272	&38	&29.23	&104\\
    K417T	&37748	&5	&0.0116	&433	&37	&28.46	&107\\
    S477N	&32673	&6	&0.0180	&422	&0	&0.0	&650\\
    N439K	&16154	&7	&0.1792	&159	&11	&8.46	&272\\
    K417N	&8399	&8	&0.1661	&176	&53	&40.77	&61\\
    F490S	&5617	&9	&0.4406	&52	&51	&39.23	&67\\
    S494P	&5119	&10	&0.0902	&282	&62	&47.69	&46\\
    N440K	&3379	&11	&0.6161	&22	&0	&0.0	&645 \\
    E484Q	&3229	&12	&0.0057	&442	&30	&23.08	&130\\
    L452Q	&2858	&13	&0.9802	&3	&27	&20.77	&144\\
    A520S	&2727	&14	&0.1495	&199	&3	&2.31	&497\\
    N501T	&2054	&15	&0.4514	&48	&17	&13.08	&202\\
    R357K	&1973	&16	&0.1393	&208	&5	&3.85	&388\\
    A522S	&1959	&17	&0.1283	&221	&2	&1.54	&543\\
    R346K	&1686	&18	&0.1234	&229	&6	&4.62	&380\\
    V367F	&1395	&19	&0.1764	&161	&0	&0.0	&637\\
    N440S	&1361	&20	&0.1499	&197	&2	&1.54	&542\\
    P384L	&1155	&21	&0.2681	&105	&18	&13.85&199\\
    Y449S	&1146	&22	&-0.8112	&632	&85	&65.38	&16\\
    D427N	&1106	&23	&-0.1133	&558	&1	&0.77	&589\\
    R346S	&1037	&24	&0.0374	&386	&20	&15.38	&182\\
    A475V	&891	&25	&0.3069	&94	&10	&7.69	&289\\
    \bottomrule 
    \end{tabular}
    }
\end{table}

First, the 10 most observed or fast-growing RBD mutations are N501Y, L452R, T478K, E484K, K417T, S477N, N439K, K417N, F490S, and S494P, as shown in Table \ref{tab:WebMutaion}. Inclusively, these top mutations strengthen their BFEs and become more infectious, following the natural selection mechanism \cite{chen2020mutations}.    Figure \ref{fig:6M0J}{\bf b} shows that the frequencies of the top three mutations increased dramatically since 2021 due to Alpha, Beta, Gamma, Delta, and other variants. 
Second, among the top 25 most observed RBD mutations, 
T478K, 
L452Q
N440K, 
L452R, 
N501Y, 
N501T, 
F490S, 
A475V, and 
P384L are the 8  most infectious ones judged by their ability to strengthen the binding with ACE2, as shown in Figure \ref{fig:6M0J}{\bf c} .  
The BFE changes of S protein and ACE2 for mutation T478K is nearly 1.00 kcal/mol, which strongly enhances the binding of the RBD–ACE2 complex \cite{cherian2021sars}. Together with L452R (BFE change: 0.58kcal/mol), T478K makes Delta the most infectious variant in VOCs.  
Third, among the top 25 most observed RBD  mutations,
Y449S,
S494P,	 
K417N,
F490S,
L452R,
E484K,
K417T,
E484Q, 
L452Q, and 
N501Y are the 10 most antibody disruptive ones, judged by their interactions with 130 antibodies shown in Figure \ref{fig:6M0J}{\bf c}. It can be seen that mutations L452R, E484K, K417T, K417N, F490S, and S494P disrupt more than 30\% of antibody-RBD complexes,  while mutations E484K and K417T may disrupt nearly 30\% antibody-RBD complexes, indicating their  disruptive ability to the efficacy and reliability of antibody therapies and vaccines. The most dangerous mutations are the ones that are both infectivity-strengthening and antibody disruptive. Four RBD mutations, N501Y, L452R, F490S,  and L452Q, appear in both lists and are key mutations in WHO's VOC and VOI lists.  Among them, F490S  and L452Q are the key RBD mutations in Lambda, making Lambda a more dangerous emerging variant than Delta. Note that high-frequency mutation S477N does not significantly weaken any antibody and RBD binding, and thus does not appear in any prevailing variants.

\subsection{Vaccine-breakthrough S protein RBD co-mutations}


\begin{figure}[ht!]
	\centering
	\includegraphics[width = 1.0\textwidth]{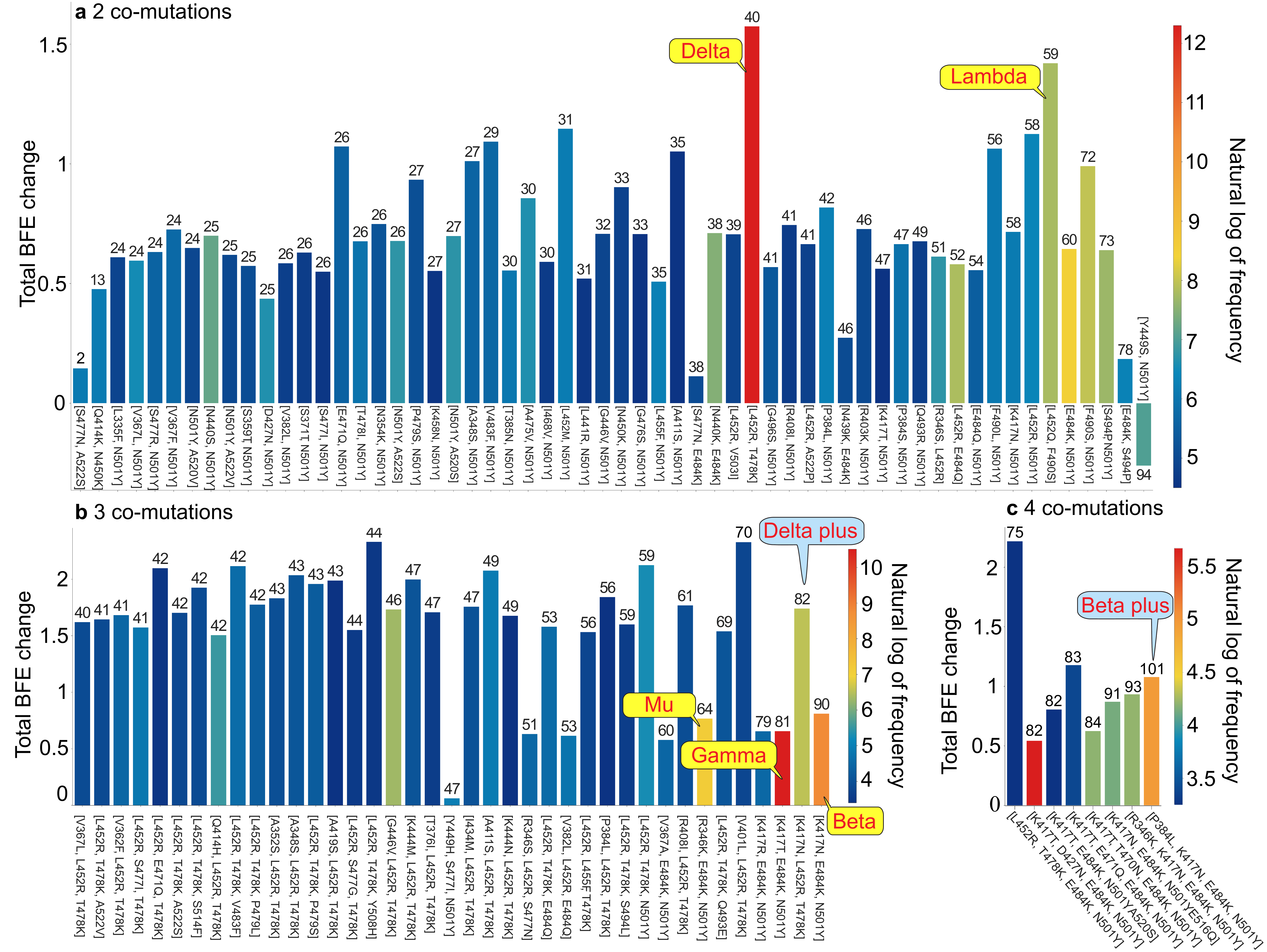}
	\caption{Properties of RBD co-mutations.	{\bf a} Illustration of RBD 2 co-mutations with a frequency greater than 90. {\bf b} Illustration of RBD 3 co-mutations with a frequency greater than 30. {\bf c} Illustration of RBD 2 co-mutations with a frequency greater than 20. Here, the $x$-axis lists  RBD co-mutations and the $y$-axis represents the predicted total BFE change of each set of RBD co-mutations. The number on the top of each bar is the AI-predicted number of antibody and RBD complexes that may be significantly disrupted by the set of RBD co-mutations, and the color of each bar represents the natural log of frequency for each set of RBD co-mutations. (Please check the interactive HTML files in the Supporting Information S2.2.4 for a better view of these plots.)}
	\label{fig:barplot}
\end{figure}

The recent surge in COVID-19 infections is due to the occurrence of RBD co-mutations that combine two or more infectivity-strengthening mutations. The most dangerous future SARS-CoV-2 variants must be RBD co-mutations that combine infectivity-strengthening mutation(s) with antibody disruptive mutation(s). A list of 1,139,244 RBD co-mutations that are decoded from 1,489,884 complete SARS-CoV-2 genome sequences can be found in Section S2.1.3 of the Supporting Information, and all of the non-degenerate RBD co-mutations with their frequencies, antibody disruption counts, total BFE changes, and the first detection dates and countries can be found in Section S2.1.4 of the Supporting Information. Figure \ref{fig:barplot} illustrates the properties of S protein RBD 2, 3, and 4 co-mutations. The height of each bar shows the predicted total BFE change of each set of co-mutations on RBD, the color represents the natural log of frequency for each set of RBD co-mutations, and the number at the top of each bar is the AI-predicted number of antibody-RBD complexes that each set of RBD co-mutations may disrupt based on a total of 130 RBD and antibody complexes. Notably, for a specific set of co-mutations, the higher the number at the top of the bar is, the stronger ability to break through vaccines will be. From Figure \ref{fig:barplot},  RBD 2 co-mutation set [L452R, T478K] (Delta variant) has the highest frequency (219,362) and the highest  BFE change (1.575 kcal/mol). Moreover, the Delta variant would disrupt 40 antibody-RBD complexes, suggesting that Delta would not only enhance the infectivity but also be a vaccine breakthrough variant. Moreover, [L452Q, F490S] (Lambda) is another co-mutation with high frequency, high BFE changes (1.421 kcal/mol), and high antibody disruption count (59). In addition, Lambda is considered to be more dangerous than Delta due to its higher antibody disruption count. Further, [R346K, E484K, N501Y] (Mu variant) has a BFE change of 0.768 kcal/mol and  high antibody disruption count (60). It is not as infectious as Delta and Lambda, but has a similar ability as Lambda in escaping vaccines. Note that among all VOCs and VOIs, Beta has the highest ability to break through vaccines, but its infectivity is relatively low (BFE change: 0.656 kcal/mol). 
Furthermore, high-frequency 2 co-mutation sets [E484K, N501Y], [F490S, N501Y], and [S494P, N501Y] are all considered to be the emerging variants that have the potential to escape vaccines. From Figure \ref{fig:barplot}, three 3 co-mutation sets [R345K, E484K, N501Y] (Mu), [K417T, E484K, N501Y] (Gamma),  and [K417N, E484K, N501Y] (Beta) draw our attention. They are all the prevailing three co-mutations with moderate BFE changes but very high antibody disruption count (more than 60). With a BFE change of 1.4 kcal/mol and antibody disruption count of 82, co-mutation set [K417N, L452R, T478K] (Delta plus) appears to be more dangerous than all of the current VOCs and VOIs. 
For 4 co-mutations in Figure \ref{fig:barplot} {\bf c}, [P384L, K417N, E484K, N501Y] (Beta plus) could penetrate all vaccines due to its highest antibody disruption count of 101. We would like to address that all of the co-mutations sets, except for  [Y449S, N501Y] in \autoref{fig:barplot} have positive BFE changes, following the natural selection. 
We anticipate that although co-mutation sets [V401L, L452R, T478K], [L452R, T478K, N501Y], [A411S, L452R, T478K], and [L452R, T478K, E484K, N501Y] have relatively low frequencies at this point, they may become dangerous variants soon due to their large BFE changes and antibody disruption counts. 

\begin{figure}[ht!]
	\centering
	\includegraphics[width = 0.6\textwidth]{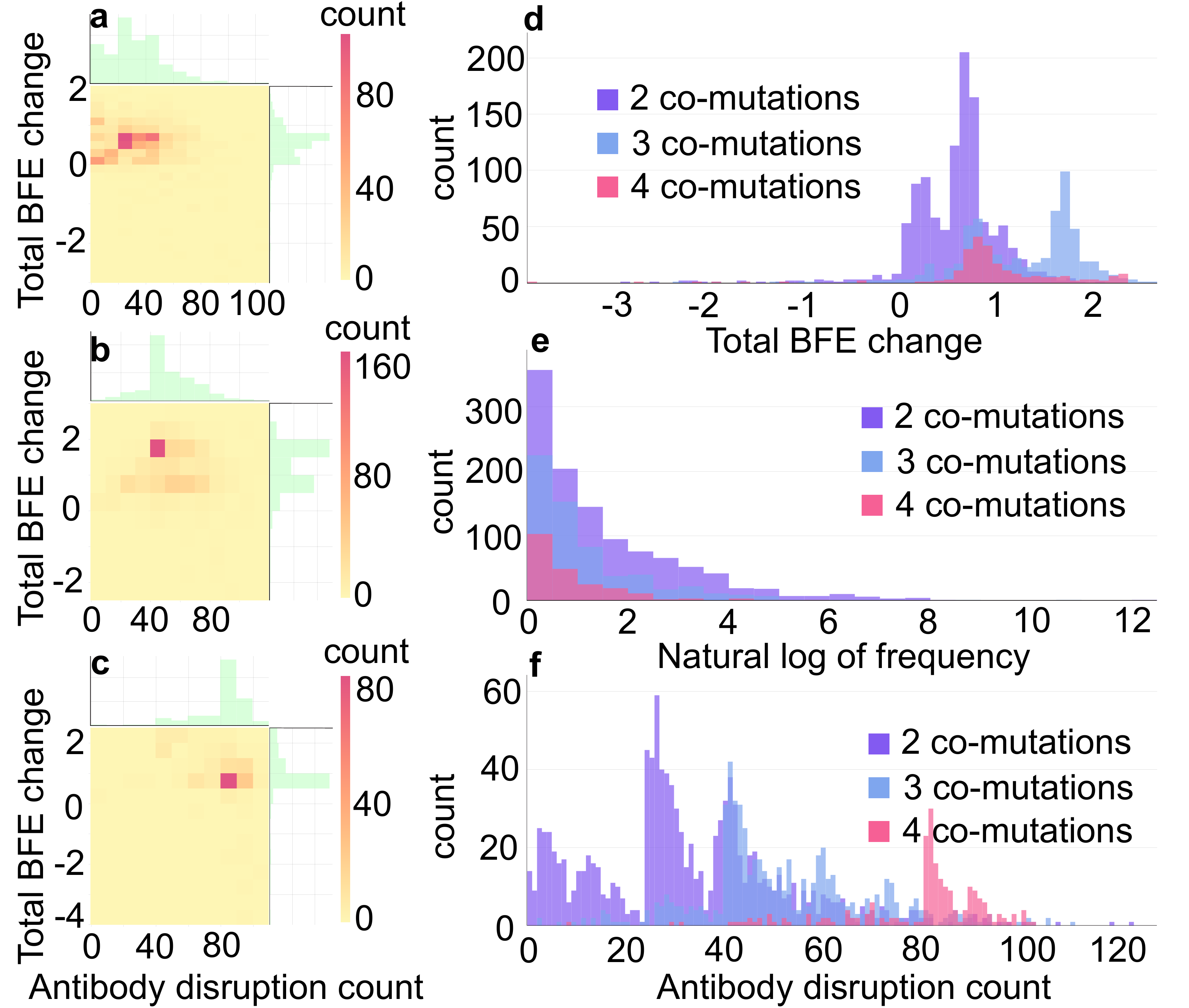}
	\caption{{\bf a} 2D histograms of antibody disruption count and total BFE changes for RBD 2 co-mutations. {\bf b} 2D histograms of antibody disruption count and total BFE changes for RBD 3 co-mutations. {\bf c} 2D histograms of antibody disruption count and total BFE changes for RBD 4 co-mutations. {\bf d} The histograms of total BFE changes for RBD co-mutations. {\bf e} The histograms of the natural log of frequency for RBD co-mutations. {\bf f} The histograms of antibody disruption count for RBD co-mutations. In figures {\bf a}, {\bf b}, and {\bf c}, the color bar represents the number of co-mutations that fall into the restriction of $x$-axis and $y$-axis. The reader is referred to the web version of these plots in the Supporting Information S2.2.2 and S2.2.3.}
	\label{fig:distribution}
\end{figure}

It is important to understand the general trend of SARS-CoV-2 evolution. 
To this end, we carry out the statistical analysis of RBD co-mutations.  
Among 1,489,884 SARS-CoV-2 genome isolates, a total of 1,113 distinctive 2 co-mutations, 612 distinctive 3 co-mutations, and 217 distinctive 4 co-mutations are found. 
Figures  \ref{fig:distribution} {\bf a}, {\bf b}, and {\bf c} illustrate the 2D histograms of 2, 3, and 4 co-mutations, respectively. The $x$-axis is the number of antibody disruption counts, and the $y$-axis shows the total BFE change. Figure \ref{fig:distribution} {\bf a} shows that there are 82 RBD 2 co-mutations that have BFE changes in the range of [0.600, 0.799] kcal/mol and will disruptive 40 to 49 antibodies. According to Figure \ref{fig:distribution} {\bf b}, there are 170 unique 3 co-mutations that have large BFE changes of S protein and ACE2 in the range of [1.500, 1.999] kcal/mol. In \autoref{fig:distribution} {\bf c}, it is seen that almost all of the 4 co-mutations on RBD have the BFE changes greater than 0.5 kcal/mol and weaken the binding of S protein with at least 60 antibodies. 
Figures \ref{fig:distribution}{\bf d}, {\bf e}, and {\bf f} are the histograms of total BFE changes, natural log of frequencies, and antibody disruption counts for RBD 2, 3, and 4 co-mutations. It can be found that most of the 2, 3, and 4 RBD co-mutations have positive total BFE changes, and the larger number of RBD co-mutations is, the higher number of antibody disruption count will be. In summary, co-mutations with a larger number of antibody disruptive counts and high BFE changes will grow faster. We anticipate that when most of the population is vaccinated, vaccine-resistant mutations will become a more viable mechanism for viral evolution.     

\subsection{Emerging breakthrough variants in COVID-19 devastated countries}

\begin{figure}[ht!]
	\centering
	\includegraphics[width = 0.95\textwidth]{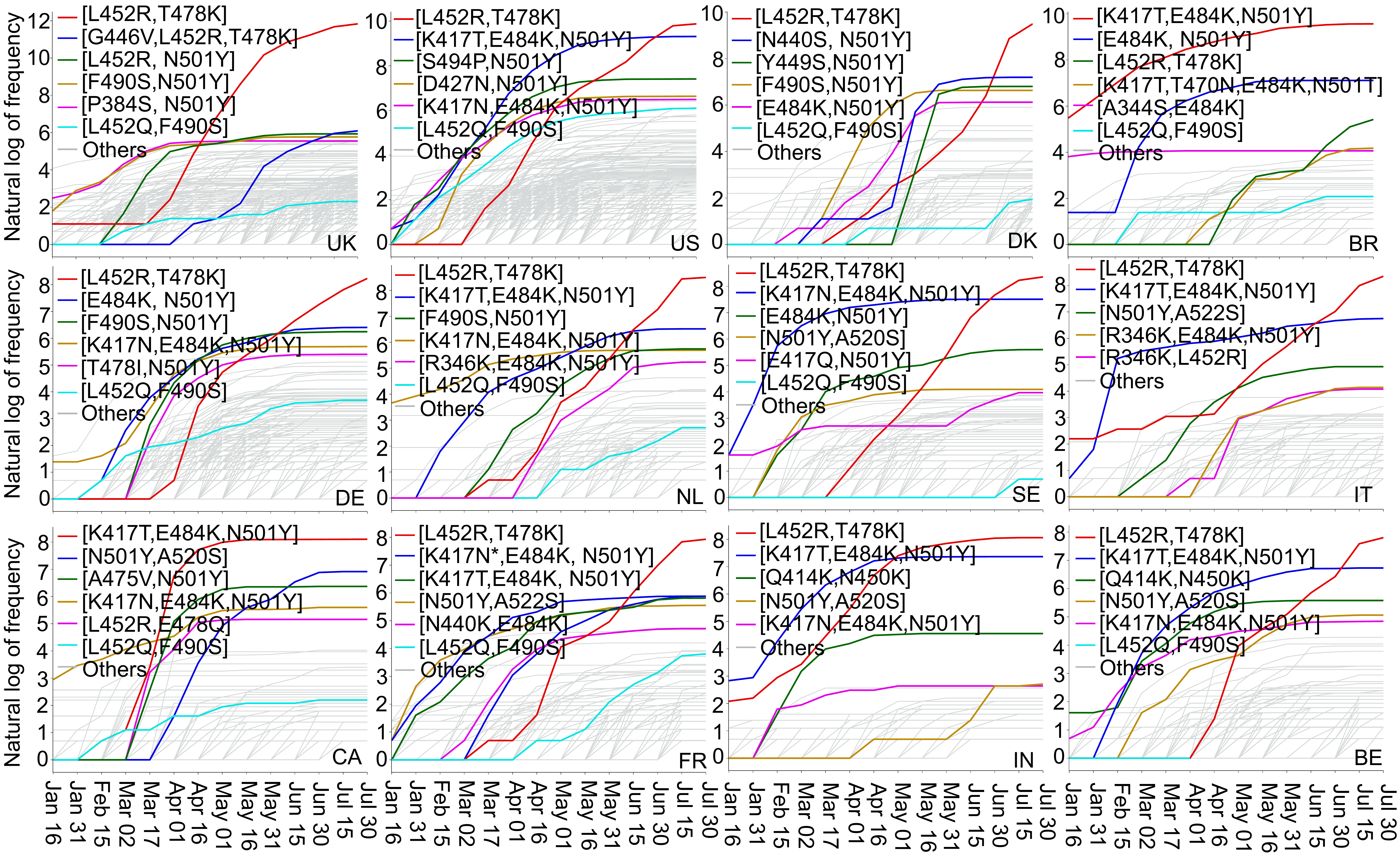}
	\caption{ Illustration of the time evolution of 2, 3, and 4 co-mutations on the S protein RBD of SARS-CoV-2 from January 01, 2021, to July 31, 2021, in 12 COVID-19 devastated countries: the United Kingdom (UK), the United States (US), Denmark (DK), Brazil (BR), Germany (DE), Netherlands (NL), Sweden (SE), Italy (IT), Canada (CA), France (FR), India (IN), and Belgium (BE). The $y$-axis represents the natural log frequency of each RBD co-mutation. The top 5 high-frequency co-mutations in each country are marked by red, blue, green, yellow, and pink lines. The cyan line is for the RBD co-mutation [L452Q, F490S] on the Lambda variant, and the other co-mutations are marked by light grey lines. Notably, there are two blues lines in the panel of FR due to the same frequency of [K417N, E484K, N501Y] and [E484K, N501Y]. (Please check the interactive HTML files in the Supporting Information S2.2.1 for a better view of these plots.)}
	\label{fig:lineplot}
\end{figure}


Our analysis of RBD mutations reveals the recent global surge of infections due to RBD co-mutations. However, due to the difference in the rate of vaccination, COVID-19 control and prevention measures, medical infrastructure, population structures, etc., each country may have a different pattern of RBD co-mutations and follow a different trajectory of SARS-CoV-2 transmission and evolution. Therefore, we analyze the RBD 2, 3, and 4 co-mutations in 20 countries that have the high frequency of SARS-CoV-2 genome isolates, including the United Kingdom (UK), the United States (US), Denmark (DK), Brazil (BR), Germany (DE), Netherlands (NL), Sweden (SE), Italy (IT), Canada (CA), France (FR), India (IN), and Belgium (BE), as well as  Ireland (IE), Spain (ES), Chile (CL), Portugal (PT), Mexico (MX), Singapore (SG), Turkey (TR), and Finland (FL). Figure \ref{fig:lineplot} shows the time evolution of 2, 3, and 4 co-mutations on the S protein RBD of SARS-CoV-2 from January 01, 2021, to July 31, 2021, in 12 COVID-19 devastated countries. The plots of the other 8 countries can be found in the Supporting Information S3. The top 5 high-frequency co-mutations in each country are marked by red, blue, green, yellow, and pink lines. The cyan line is for the RBD co-mutation set [L452Q, F490S] on the Lambda variant, which is more penetrative to vaccines than the Delta.  Light grey lines mark the other co-mutations. The RBD co-mutation set [L452R, T478K] (Delta) with 1.575 kcal/mol BFE change was first found in IN in early January 2021, and the number of this variant increases rapidly around the world in a short period. Later on, in early March 2021, the UK, US, DK, DE, NL, SE, IT, FR, BE reported the appearance of [L452R, T478K] in early March 2021, and eventually [L452R, T478K] became a dominated co-mutation, which is consistent to the finding that Delta variant remains largely susceptible to infection. The co-mutation set [K417T, E484K, N501Y] (Gamma) with BFE change of 0.656 kcal/mol was first found in Brazil in early January 2021, and then it became the most dominated co-mutation in Brazil and Canada, and the second dominated co-mutation in the US, NL, SE, IT, FR, IN, and BE. Notably, co-mutaion set [G446V, L452R, T478K] in the UK with BFE change of 1.733 kcal/mol and 46 antibody disruption counts appears to be a dangerous set of co-mutations that may affect the infectivity and vaccine/antibodies efficacy shortly. Moreover, co-mutation set [N501Y, A520S] has quickly increased IN and BE since April 16, 2021. Considering the BFE change and antibody disruptive count of co-mutation set [N501Y, A520S]  is 0.699 and 27, we suggest monitoring this variant in IN and BE. Furthermore, the co-mutation set [K417N, T470N, E484K, N501T] that was first found in BR on April 06, 2020, has a BFE change of 0.625 kcal/mol and antibody disruption count 84, is an emerging vaccine breakthrough co-mutation in Brazil. In addition, co-mutation set [L452Q, F490S] (cyan lines) on Lambda variant was recently drawing much attention due to its potential ability to resist vaccines and enhance the infectivity, which is consistent with our predictions that co-mutation set [L452Q, F490S] has a relatively significant BFE change of S protein and ACE2 (1.421kcal/mol) and would reduce the RBD binding with 59 antibodies. Lambda has already spread out in every country in Figure \ref{fig:lineplot}.

\section{Methodology and validation}
\begin{figure}[ht!]
	\setlength{\unitlength}{1cm}
	\begin{center}
		\includegraphics[width = 1\textwidth]{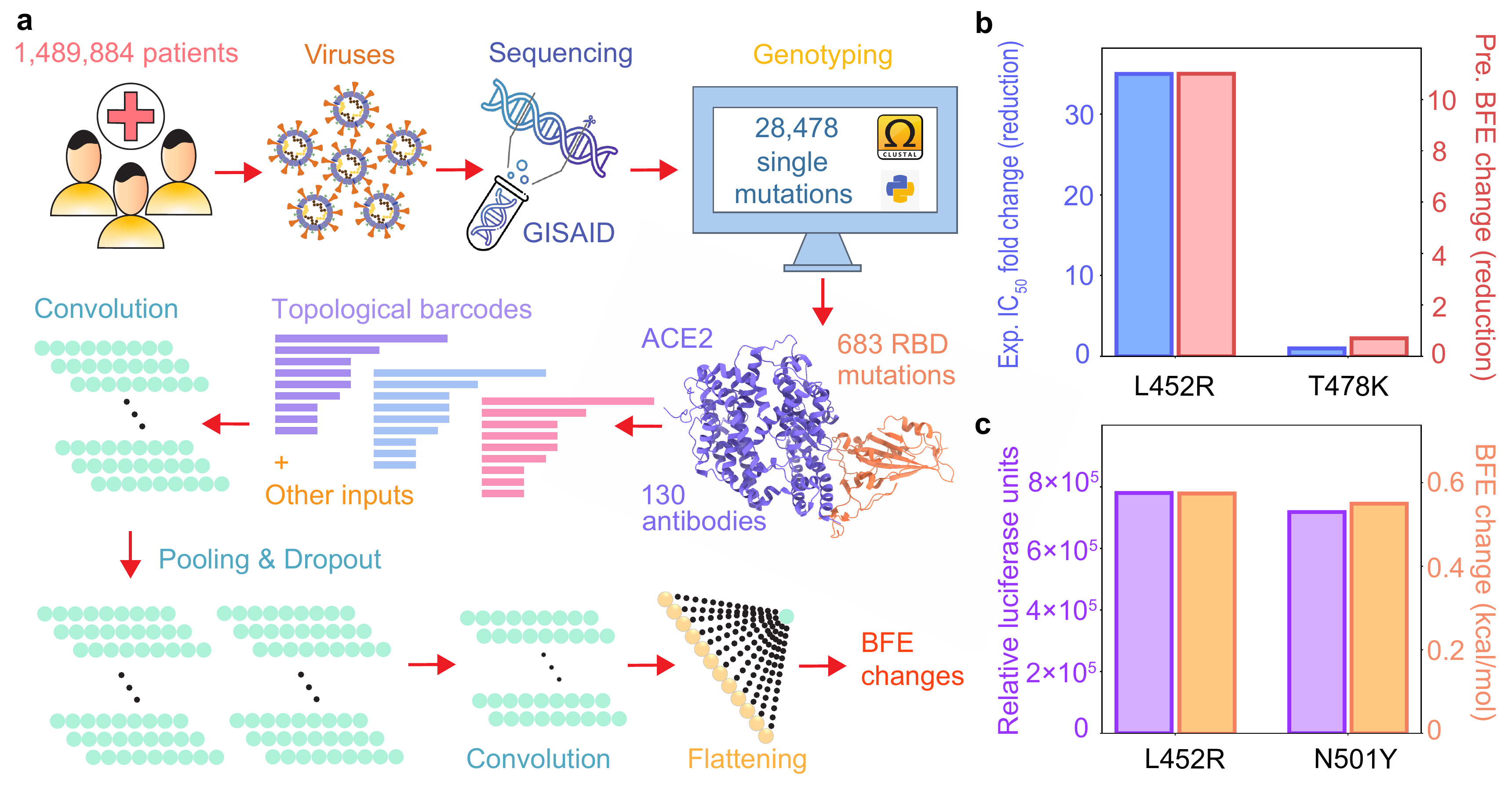}
		\caption{{\bf a} Illustration of genome sequence data pre-processing and BFE change predictions. {\bf b} Comparison of experimental CT-P59 IC$_{50}$ fold change (reduction) and predicted BFE changes induced by mutations L452R and T478K. {\bf c} Comparison of predicted BFE changes and relative luciferase units \cite{deng2021transmission} for pseudovirus infection changes of ACE2 and S protein complex induced by mutations L452R and N501Y.}
		\label{fig_combine_all}
	\end{center}
\end{figure}

In this section, the work flow of deep learning-based BFE change predictions of protein-protein interactions induced by mutations for the present SARS-CoV-2 variant analysis and prediction will be firstly introduced, which includes four steps as shown in \autoref{fig_combine_all}: (1) Data pre-processing; (2) training data preparation; (3) feature generations of protein-protein interaction complexes; (4) prediction of protein-protein interactions by deep neural networks (check Section S5 in Supporting information). Next, the validation of our machine learning-based model will be demonstrated, suggesting consistent and reliable results compared to the experimental deep mutations data.

\subsection{Data pre-processing and SNP genotyping}
The first step is to pre-process the original SARS-CoV-2 sequences data. In this step, a total of 1,489,884 complete SARS-CoV-2 genome sequences with high coverage and exact collection date are downloaded from the GISAID database \cite{shu2017gisaid} (\url{https://www.gisaid.org/}) as of August 05, 2021. Complete SARS-CoV-2 genome sequences are available from the GISAID database \cite{shu2017gisaid}.  Next, the 1,489,884 complete SARS-CoV-2 genome sequences were rearranged according to the reference genome downloaded from the GenBank (NC\_045512.2)\cite{wu2020new}, and multiple sequence alignment (MSA) is applied by using Cluster Omega with default parameters. Then, single nucleotide polymorphism (SNP) genotyping is applied to measure the genetic variations between different isolates of SARS-CoV-2 by analyzing the rearranged sequences \cite{yin2020genotyping,kim2007snp}, which is be of paramount importance for tracking the genotype changes during the pandemic. The SNP genotyping captures all of the differences between patients' sequences and the reference genome, which decodes a total of 28,478 unique single mutations from 1,489,884 complete SARS-CoV-2 genome sequences. Among them, 4,653 non-degenerate mutations on S protein and 683 non-degenerate mutations on the S protein RBD (S protein residues from 329 to 530) are detected. In this work, the co-mutation analysis is more crucial than the unique single mutation analysis. Therefore, for each SARS-CoV-2 isolate, we extract the all of the mutations on S protein RBD, which is called a RBD co-mutation for a specific isolates. By doing this, a total of 1,139,244 RBD co-mutations are captured. Notably, the SARS-CoV-2 unique single mutations in the world is available at \href{https://users.math.msu.edu/users/weig/SARS-CoV-2_Mutation_Tracker.html}{Mutation Tracker}. The analysis of RBD mutations is available at \href{https://weilab.math.msu.edu/MutationAnalyzer/}{Mutation Analyzer}.

\subsection{Methods for BFE change predictions}
In this section, the process of the machine learning-based BFE change predictions is introduced. Once the data pre-processing and SNP genotyping is carried out, we will firstly proceed with the training data preparation process, which plays a key role in reliability and accuracy. A library of 130 antibodies and RBD complexes as well as an ACE2-RBD complex are obtained from Protein Data Bank (PDB). RBD mutation-induced BFE changes of these complexes are evaluated by the following machine learning model. According to the emergency and the rapid change of RNA virus, it is rare to have massive experimental BFE change data of SARS-CoV-2, while, on the other hand, next-generation sequencing data is relatively easy to collect. In the training process, the dataset of BFE changes induced by mutations of the SKEMPI 2.0 dataset \cite{jankauskaite2019skempi} is used as the basic training set, while next-generation sequencing datasets are added as assistant training sets. The SKEMPI 2.0 contains 7,085 single- and multi-point mutations and 4,169 elements of that in 319 different protein complexes used for the machine learning model training. The mutational scanning data consists of experimental data of the binding of ACE2 and RBD induced mutations on ACE2\cite{procko2020sequence} and RBD\cite{starr2020deep, linsky2020novo}, and the binding of CTC-445.2 and RBD with mutations on both protein\cite{linsky2020novo}.

Next, the feature generations of protein-protein interaction complexes is performed. The element-specific algebraic topological analysis on complex structures is implemented to generate topological bar codes \cite{chen2021revealing, carlsson2009topology,edelsbrunner2000topological,xia2014persistent}. In addition, biochemistry and biophysics features such as Coulomb interactions, surface areas, electrostatics, et al., are combined with topological features \cite{chen2021prediction}. The detailed information about the topology-based models will be demonstrated in \autoref{sec:topology}. Lastly, deep neural networks for SARS-CoV-2 are constructed for the BFE change prediction of protein-protein interactions \cite{chen2021revealing}. The detailed descriptions of dataset and machine learning model are found in the literature \cite{chen2020mutations,wang2020mutations,chen2021revealing} and are available at \href{https://github.com/WeilabMSU/TopNetmAb}{TopNetmAb}.

\subsection{Feature generation for machine learning model}\label{sec:topology}
Among all features generated for machine learning prediction, the application of topology theory makes the model to a whole new level. Those summarized as other inputs are called as auxiliary features and are described in Section S4 of the Supporting Information. In this section, a brief introduction about the theory of topology will be discussed. Algebraic topology \cite{carlsson2009topology,edelsbrunner2000topological} has achieved tremendous success in many fields including biochemical and biophysical properties\cite{xia2014persistent}. Special treatment should be implemented for biology applications to describe element types and amino acids in poly-peptide mathematically, which element-specific and site-specific persistent homology \cite{wang2020topology,chen2020mutations}. To construct the algebraic topological features on protein-protein interaction model, a series of element subsets for complex structures should be defined, which considers atoms from the mutation sites, atoms in the neighborhood of the mutation site within a certain distance, atoms from antibody binding site, atoms from antigen binding site, and atoms in the system that belong to type of \{C, N, O\}, $\mathcal{A}_\text{ele}(\text{E})$. Under the element/site-specific construction, simplicial complexes is constructed on point clouds formed by atoms. For example, a set of independent $k\!+\!1$ points is from one element/site-specific set $U=\{u_0, u_1, ...,u_k\}$. The $k$-simplex $\sigma$ is a convex hull of $k\!+\!1$ independent points $U$, which is a convex combination of independent points. For example, a $0$-simplex is a point and a $1$-simplex is an edge. Thus, a $m$-face of the $k$-simplex with $m\!+\!1$ vertices forms a convex hull in a lower dimension $m<k$ and is a subset of the $k\!+\!1$ vertices of a $k$-simplex, so that  a sum of all its $(k\!-\!1)$--faces is the boundary of a $k$--simplex $\sigma$ as
\begin{equation}
\partial_k\sigma = \sum_{i=1}^{k}(-1)^i\langle u_0, ..., \hat{u}_i, ..., u_k\rangle ,
\label{eq_boundary_operator}
\end{equation}
where $\langle u_0, ..., \hat{u}_i, ..., u_k\rangle $ consists of all vertices of $\sigma$ excluding $u_i$. The collection of finitely many simplices is a simplicial complex. In the model, the Vietoris-Rips (VR) complex (if and only if $\mathbb{B}(u_{i_j}, r)\cap\mathbb{B}(u_{i_{j'}}, r)\ne\emptyset$ for ${j}, {j'} \in [0,k]$) is for dimension 0 topology, and alpha complex (if and only if $\cap_{u_{i_j}\in\sigma}\mathbb{B}(u_{i_j}, r)\ne\emptyset$) is for point cloud of dimensions 1 and 2 topology \cite{xia2014persistent}. 

The $k$-chain $c_k$ of a simplicial complex $K$ is a formal sum of the $k$-simplices in $K$, which is $c_k=\sum\alpha_i\sigma_i$, where  $\alpha_i$ is coefficients and is chosen to be $\mathbb{Z}_2$. Thus, the boundary operator on a $k$-chain $c_k$ is
\begin{equation}
\partial_kc_k=\sum\alpha_i\partial_k\sigma_i,
\label{eq_boundary_operator_chain}
\end{equation}
such that $\partial_k:C_k\rightarrow C_{k-1}$ and follows from that boundaries are boundaryless $\partial_{k-1}\partial_k=\emptyset$. 
A chain complex is
\begin{equation}
\cdots \stackrel{\partial_{i+1}}\longrightarrow C_i(K) \stackrel{\partial_i}\longrightarrow C_{i-1}(K) \stackrel{\partial_{i-1}}\longrightarrow \cdots \stackrel{\partial_2} \longrightarrow C_{1}(K) \stackrel{\partial_{1}}\longrightarrow C_0(K) \stackrel{\partial_0} \longrightarrow 0,
\label{eq_chain_complex}
\end{equation}
as a sequence of complexes by boundary maps. Therefore, the Betti numbers are given as the ranks of $k$th homology group $H_k$ as $\beta_k=\text{rank}(H_k)$, where $H_k=Z_k/B_k$, $k$-cycle group $Z_k$ and the $k$-boundary group $B_k$. The Betti numbers are the key for topological features, where $\beta_0$ gives the number of connected components, such as number of atoms, $\beta_1$ is the number of cycles in the complex structure, and $\beta_2$ illustrates the number of cavities. This presents abstract properties of the 3D structure.

Finally, only one simplicial complex couldn't give the whole picture of the protein-protein interaction structure. A filtration of a topology space is needed to extract more properties.
A filtration is a nested sequence such that
\begin{equation}
\emptyset = K_0 \subseteq K_1 \subseteq \cdots \subseteq K_m = K.
\label{eq_filtration}
\end{equation}
Each element of the sequence could generate the Betti numbers $\{\beta_0, \beta_1, \beta_2\}$ and consequentially, a series of Betti numbers in three dimensions is constructed and applied to be the topological fingerprints in Figure~\ref{fig_combine_all}{\bf a}.

\subsection{Validation}
The validation of our machine learning predictions for mutation-induced BFE changes compared to experimental data has been demonstrated in recently published papers \cite{chen2021prediction,chen2021revealing}. Firstly, we showed high correlations of experimental deep mutational enrichment data and predictions for the binding complex of SARS-CoV-2 S protein RBD and protein CTC-445.2 \cite{chen2021prediction} and the binding complex of SARS-CoV-2 RBD and ACE2 \cite{chen2021revealing}. In comparison with experimental data on antibody therapies in clinical trials of emerging mutations, our predictions achieve a Pearson correlation at 0.80 \cite{chen2021revealing}. Considering the BFE changes induced by RBD mutations for ACE2 and RBD complex, predictions on mutations L452R and N501Y have a highly similar trend with experimental data \cite{chen2021revealing}. Meanwhile, as we presented in \cite{wang2021vaccine}, high-frequency mutations are all having positive BFE changes. Moreover, for multi-mutation tests, our BFE change predictions have the same pattern with experimental data of the impact of SARS-CoV-2 variants on major antibody therapeutic candidates, where the BFE changes are accumulative for co-mutations \cite{chen2021revealing}.

Recent studies on potency of mAb CT-P59 in vitro and in vivo against Delta variants\cite{lee2021therapeutic} show that the neutralization of CT-P59 is reduced by   L452R (13.22 ng/mL) and is retained against T478K (0.213 ng/mL). In our predictions \cite{chen2021revealing}, L452R induces a negative BFE change (-2.39 kcal/mol), and T478K produces a positive BFE change (0.36 kcal/mol). In Figure~\ref{fig_combine_all}{\bf b}, the fold changes for experimental and predicted values are presented.  
Additional, in Figure~\ref{fig_combine_all}{\bf c}, a comparison of the experimental pseudovirus infection changes and predicted BFE change of ACE2 and S protein complex induced by mutations L452R and N501Y, where the experimental data is obtained in a reference to D614G and reported in relative luciferase units \cite{deng2021transmission}. It indicates that the binding of RBD and ACE2 dominates the infectivity of SARS-CoV-2. More details can be found in Section S6 of Supporting information.

\section*{Data and model availability}
The SARS-CoV-2 SNP data in the world is available at \href{https://users.math.msu.edu/users/weig/SARS-CoV-2_Mutation_Tracker.html}{Mutation Tracker}. The most observed SARS-CoV-2 RBD mutations are available at \href{https://weilab.math.msu.edu/MutationAnalyzer/}{Mutaton Analyzer}. The information of 130 antibodies with their corresponding PDB IDs can be found in the Supplementary Data. The SARS-CoV-2 S protein RBD SNP and non-degenerate co-mutations data can be found in Section S2.1.4 of the Supporting Information. The TopNetTree model is available at \href{https://github.com/WeilabMSU/TopNetmAb}{TopNetmAb}.

\section*{Supporting information}
The supporting information is available for 

\begin{enumerate}
    \item[S1] Overview of SARS-CoV-2 prevailing and emerging variants
    \item[S2] Supplementary data: The Supplementary\_Data.zip contains two folders: S2.1: {\color{teal} Excel folder}: A total of 4 files are in this folder: S2.1.1: antibodies\_disruptmutation.csv shows the name of antibodies disrupted by mutations. S2.1.2: antibodies.csv lists the PDB IDs for all of the 130 SARS-CoV-2 antibodies. S2.1.3: RBD\_comutation\_residue\_08052021.csv lists all of the SNPs of RBD co-mutations up to August 05, 2021. S2.1.4: Track\_Comutation\_08052021.xlsx preserves all of the non-degenerate RBD co-mutations with their frequencies, antibody disruption counts, total BFE changes, and the first detection dates and countries. S2.2: {\color{teal} HTML folder}: A total of 29 HTML files containing: S2.2.1: 20 HTML files for the for the time evolution of 2, 3, and 4 co-mutations on the S protein RBD of SARS-CoV-2 from January 01, 2021 to July 31, 2021, in 12 COVID-19 devastated countries. S2.2.2: Three 2D histograms are given for antibody disruption counts and total BFE changes for RBD 2 co-mutations, 3 co-mutations, and 4 co-mutations.  S2.2.4: Three histograms of total BFE changes, antibody disruption count, and natural log of frequencies for RBD 2 co-mutations, 3 co-mutations, and 4 co-mutations. S2.2.4: Three barplots for RBD 2, 3, 4 co-mutations with a frequency greater than 90, 30, and 20, respectively. 
    \item[S3] Supplementary figures: the line plot of the time evolution of 2, 3, and 4 co-mutations on the S protein RBD of SARS-CoV-2 from January 01, 2021, to July 31, 2021, in 8 COVID-19 devastated countries.
    \item[S4] Supplementary feature generation: detailed description of feature generations.
    \item[S5] Supplementary machine learning methods: detailed description of machine learning method implemented in this work
    \item[S6] Supplementary validation: validations of our machine learning predictions with experimental data.
\end{enumerate}

\section*{Acknowledgment}
This work was supported in part by NIH grant  GM126189, NSF grants DMS-2052983,  DMS-1761320, and IIS-1900473,  NASA grant 80NSSC21M0023,  Michigan Economic Development Corporation, MSU Foundation,  Bristol-Myers Squibb 65109, and Pfizer. GWW thanks the discussion with Dr. Peter Lyster which inspired this work. 

\bibliographystyle{unsrt}
\bibliography{refs}
\end{document}